\documentclass[
    ,final            
  ]
  {aipproc}

\layoutstyle{6x9}

\usepackage{epsfig}

\begin{document}

\title{
{\noindent \small UNITU-THEP-17/2004
\hspace*{1cm} IPPP/04/76  
\hspace*{1cm} DCPT/04/152 
\hfill hep-ph/0411367}\\
{~}\\
Analytic structure of\\ the Landau gauge gluon propagator}

\classification{12.38.Aw 14.65.Bt 14.70.Dj 12.38.Lg 11.30.Rd 11.15.Tk
  02.30.Rz}
\keywords      {Strong QCD, Green's functions, Confinement, 
Dyson--Schwinger equations}

\author{R.~Alkofer}
{address={Institute for Theoretical Physics, University of
  T\"ubingen, D-72076 T\"ubingen, Germany}}

\author{W.~Detmold}
{address={Department of Physics, University of Washington,
             Box  351560, Seattle WA 98195, USA}}

\author{C.~S.~Fischer}
{address={IPPP, University of Durham, Durham DH1 3LE, U.K.}}

\author{P.~Maris}
{address={U.\ of Pittsburgh, Dept.\ of Physics and Astronomy,
        100 Allen Hall, 
        Pittsburgh PA 15260, USA        }}

\begin{abstract}
The results of different non-perturbative studies agree on a power law as
the infrared behavior of the Landau gauge gluon propagator. This propagator
violates positivity and thus indicates the absence of the transverse gluons
from the physical spectrum, {\it i.e.\/} gluon confinement. A simple analytic
structure for the gluon propagator is proposed capturing all of its features. 
We comment also on related investigations for the Landau gauge quark propagator.
\end{abstract}

\maketitle

In this talk a study of the analytic properties of the gluon propagator in
Landau gauge QCD will be presented. Hereby results from
non-perturbative calculations of this propagator are employed. A detailed
account of this  investigation can be found in ref.~\cite{Alkofer:2003jj},
for three-dimensional Yang-Mills theory see also ref.\ \cite{Maas:2004se}.

In the following we will confirm previous results 
\cite{vonSmekal:1997is,Mandula:nj} on positivity violation for the gluon
propagator. We will also provide a parameterisation of the gluon propagator
that is analytic everywhere in the complex $p^2$ plane except on the real
timelike axis and  decreases to zero in every direction of the complex $p^2$
plane\footnote{We have also provided
parameterisations of the quark propagator \cite{Alkofer:2003jj}, some of which
are analytic everywhere in the complex $p^2$ plane except the timelike real
half-axis.}. Such a  behaviour satisfies the standard axioms of local quantum field 
theory except positivity.

In Landau gauge the gluon propagator can be generically written as
\begin{eqnarray}
D_{\mu \nu}(p) &=& \left(\delta_{\mu \nu} - \frac{p_\mu
      p_\nu}{p^2} \right) \frac{Z(p^2)}{p^2} \, .
  \label{gluon_prop}
\end{eqnarray}
In Euclidean quantum field theory, positivity of a propagator
can be tested by performing a Fourier transformation with respect to
Euclidean time. A violation of the condition 
\begin{eqnarray}
  \Delta(t) := \int d^3x \int \frac{d^4p}{(2\pi)^4}
  e^{i(t p_4+\vec{x}\cdot\vec{p})} \frac{Z(p^2)}{p^2}
  = \frac{1}{\pi}\int
  dp_4 \cos(t p_4) \frac{Z(p^2)}{p^2} \;\ge 0 \,,
\label{schwinger}
\end{eqnarray}
for the Schwinger function then proves violation of positivity. 
The Dyson--Schwinger equations 
(for recent reviews see {\it e.g.\/} \cite{Alkofer:2000wg}) for the ghost,
gluon and quark propagators in the Landau gauge have been solved recently in a
self-consistent truncation scheme \cite{Fischer:2003rp,Fischer:2002hn}.
Especially, one analytically obtains 
\begin{eqnarray}
  Z(p^2) \sim (p^2)^{2\kappa}, \qquad
  G(p^2) \sim (p^2)^{-\kappa},
  \label{g-power}
\end{eqnarray}
for the gluon and ghost dressing function with exponents related to
each other. In this particular truncation $\kappa$ is an irrational number,
$\kappa = (93 - \sqrt{1201})/98 \approx 0.595$  
\cite{Lerche:2002ep,Zwanziger:2001kw}. 
This result depends only slightly on the employed truncation scheme:
Infrared dominance of the gauge fixing part of the QCD action
\cite{Zwanziger:2003cf} implies infrared dominance of ghosts 
which in turn can be used to show \cite{Lerche:2002ep} that the
infrared exponents
depend only weakly on the dressing of the ghost-gluon vertex 
\cite{Schleifenbaum:2004id} and not at
all on other vertex functions \cite{Llanes-Estrada:2004jz}.
Furthermore, investigations based on the Exact Renormalisation Group Equations
 find the relations (\ref{g-power}) with an identical or
slightly lower value for $\kappa$ \cite{Pawlowski:2003hq}. 

The running coupling as it results from numerical solutions for the gluon and
ghost propagators can be accurately represented by \cite{Fischer:2003rp}
\begin{eqnarray}
\alpha_{\rm fit}(p^2) = \frac{1}{1+(p^2/\Lambda^2_{\tt QCD})}
\left( \alpha(0) + \frac{p^2}{\Lambda^2_{\tt QCD}}
\frac{4 \pi}{\beta_0}
\left(\frac{1}{\ln(p^2/\Lambda^2_{\tt QCD})}
- \frac{1}{p^2/\Lambda_{\tt QCD}^2 -1}\right)\right) . 
\label{fitB}
\end{eqnarray}
with $\beta_0=(11N_c-2N_f)/3$. 
The expression (\ref{fitB}) is analytic in the complex $p^2$ plane except the 
negative real axis $p^2<0$, {\it i.e.\/} timelike momenta, where the logarithm
produces a cut.

\begin{figure}
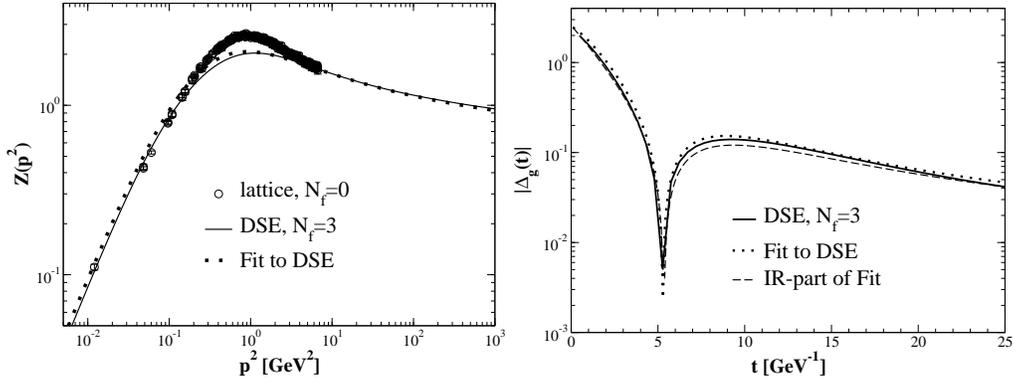

\epsfig{file=xx1.eps,width=0.45\linewidth} \hfill
\epsfig{file=xx2.eps,width=0.45\linewidth}
\caption{Left panel: Dyson--Schwinger \protect\cite{Fischer:2003rp}
and lattice results \protect\cite{Bonnet:2001uh}
for the gluon renormalization 
function $Z(p^2)$ and the fit (\protect{\ref{fitII}}).
Right panel: The results for the absolute value of the gluon 
Schwinger function from the Dyson--Schwinger solution, the fit (\protect{\ref{fitII}})
and the infrared part of this fit.}
\label{XX1}
\end{figure}

The fact that the exponent $\kappa$ in eq. (\ref{g-power}) is an irrational number 
has an important consequence: the gluon propagator possesses a cut on the 
negative real axis. It is possible to fit the non-perturbative solution 
for the gluon propagator very well without introducing further singularities. 
The fit to the gluon renormalization function
\begin{equation}
Z_{\rm fit}(p^2) = w \left(\frac{p^2}{\Lambda^2_{\tt QCD}+p^2}\right)^{2 \kappa}
 \left( \alpha_{\rm fit}(p^2) \right)^{-\gamma}
 \label{fitII}
\end{equation}
with $w= 2.65$ and $\Lambda_{\tt QCD}=520$ MeV 
is shown in fig.\ \ref{XX1}. Similar parametrizations have been explored in 
ref.\ \cite{Alkofer:2003jj}.
Hereby $w$ is a normalization parameter, and 
$\gamma $ is the one-loop value for 
the anomalous dimension of the gluon propagator. 

The  Schwinger function $\Delta (t)$ based on the fit (\ref{fitII}) is
compared to the one of the numerical DSE solution in fig.\ \ref{XX1}. To enable a
logarithmic scale the absolute value is displayed.   $\Delta (t)$ has a zero
for $t\approx 5$/GeV and is negative for larger Euclidean times: One clearly
observes positivity violations in the  gluon propagator.  The overall magnitude
$w$ is arbitrary due to renormalization  properties.   The infrared exponent
$\kappa$ is determined analytically,  and for the gluon anomalous dimension
$\gamma$ the one-loop value is used. Thus the parameterization of the gluon
propagator  has effectively only one parameter, the scale $\Lambda_{\tt QCD}$.
This and the relatively simple analytic structure gives us confidence that  
the  important features of the Landau gauge gluon propagator are given by 
(\ref{fitII}).

Finally, we want to mention that the regular infrared behaviour of the quark
propagator found from the Dyson--Schwinger equations
\cite{Fischer:2003rp} and on the lattice
\cite{Zhang:2003fa}  complicates the issue of
determining the analytic properties of the quark propagator. 
Nevertheless there is some 
evidence that the Schwinger functions related to the quark
propagator are positive definite. {\it E.g.\/}
the quark Schwinger functions can be described  accurately by a cut on the
negative real axis to the left of a singular  point at $p^2=-m^2$ where 
$m= 350 \ldots 390$ MeV 
might be attributed the meaning of an infrared constituent mass.

\begin{theacknowledgments}
RA thanks the organizers of 
{\it Quark Confinement and the Hadron Spectrum VI\/} 
for making this extraordinary conference possible.\\
This work has been supported by  a
grant from the Ministry of Science, Research and the Arts of
Baden-W\"urttemberg (Az: 24-7532.23-19-18/1 and 24-7532.23-19-18/2)
and the Deutsche For\-schungsgemeinschaft (DFG) under contract Fi 970/2-1.
\end{theacknowledgments}

\end{document}